\documentstyle[aps,epsfig]{revtex}
\begin{document}
\draft
\title{Collective excitation frequencies and vortices of a
Bose-Einstein condensed state with gravity-like interatomic attraction}

\author{Tarun Kanti Ghosh }
\address
{The Institute of Mathematical Sciences, C. I. T. Campus, Taramani, Chennai-
600 113, India.}
\date{\today}
\maketitle

\begin{abstract}
We study the collective excitations of a neutral atomic Bose-Einstein condensate with
gravity-like $1/r $ interatomic attraction induced by electromagnetic wave.
Using the time-dependent variational approach, we derive an analytical 
 spectrum for monopole and quadrupole mode frequencies of a gravity-like 
self-bound Bose condensed state at zero temperature. We also analyze the excitation 
frequencies of the Thomas-Fermi-gravity (TF-G) and gravity (G) regimes. Our result 
agrees excellently with that of Giovanazzi {\em et al}.
[ Europhysics Letters , {\bf 56}, 1 (2001)], 
which is obtained within the sum-rule approach. We also consider the vortex state. We estimate
the superfluid coherence length and the critical angular frequencies to create
a vortex around the $ z $-axis. We find that the TF-G regime can exhibit the superfluid
properties more prominently than the G-regime. We find that the monopole
mode frequency of the condensate decreases due to the presence 
of a vortex.

\end{abstract}
\noindent
\section{Introduction}

The discovery of Bose-Einstein condensation (BEC)  in a dilute alkali atomic gas opens
up new perspective in the field of many body physics \cite{anderson,web}. Most of the 
properties of these dilute gas can be explained by considering only 
two-body short range interaction which is characterized by the s-wave scattering
length $ a $\cite{web,dal}. Recently, a new kind of atomic BEC has been proposed by 
D. O'Dell {\em et al}. 
\cite{dell}. They have shown that the particular configuration of intense 
off-resonant laser beams gives rise to an effective gravity-like $ 1/r $ 
interatomic attraction between neutral atoms located well within the laser 
wavelength. This long range interaction potential is of the form,
$ U(r) = - u/r $,
where  $ u = (11 \pi/15)( I \alpha_0^2 / c \epsilon_{0}^2 \lambda_L^2) $.
 $ I $ and $ \lambda_L $ are the total laser intensity and wave length respectively.
 $ \alpha_0 $ is the atomic polarizability at the frequency $ 2 \pi c/ \lambda_L $.
In this system, the gravity-like $ 1/r $ attraction balances the pressure
due to the zero point kinetic energy and  the short range interaction potential.
 For strong induced gravity-like potential, the BEC becomes stable even without 
the external trap potential \cite{kuri}. 
 There is a competition between 
the gravity-like potential either with  the kinetic energy (G) or the two-body short range 
interatomic interaction potential characterized by the $s$-wave
scattering length $ a $ (TF-G), which gives two new regimes for new atomic BEC.
These two regimes are obtained based on the Gaussian variational ansatz 
for the ground state wave function \cite{dell}.

In the TF-G regime, collective excitation frequencies 
has been calculated numerically by solving the equations of collisionless 
hydrodynamics \cite{maz}. Moreover, in this regime, an analytic expression of 
the ground state density is obtained \cite{dell}. Within the 
sum-rule approach, 
collective excitation frequencies of a gravity-like self-bound  Bose gas has 
been discussed in Ref. \cite{gio}.
There has been no systematic and detail study on the collective excitation
frequencies and vortices of a gravitationally self-bound Bose gas by using 
the time dependent variational approach. The purpose of this paper
is to give an analytic expressions for collective excitation frequencies,  superfluid 
coherence length and  critical angular frequencies required 
to create a vortex  of a rotating Bose condensed state and to compare 
qualitatively the
results of the TF-G regime with the results obtained in the TF regime 
of an ordinary atomic BEC.  

Here, by using the time-dependent variational method, we obtain 
the excitation spectrum of a gravity-like self-bound Bose gas.
We also calculate the lower bound of the ground state energy,
monopole and quadrupole mode 
frequencies of the TF-G and the G regimes.  
Our result agrees excellently with that of ref. \cite{gio}
which is obtained within the sum-rule approach.
Next, we consider a rotating
Bose condensate state with a single vortex along the $ z $-axis.
We estimate the superfluid coherence length and the critical angular frequencies
required to create a vortex along the $ z $-axis. We find that the 
TF-G regime of a gravitationally self-bound Bose condensed state should
exhibit superfluid properties prominently than the G-regime. 
We find that the monopole mode frequency of the condensate decreases due 
to presence of the vortex.

\section{Time-dependent variational analysis }
The equation of motion of the condensate wave function is described by the generalized 
Gross-Pitaevskii equation \cite{gross},

\begin{equation}
 i \hbar \frac{\partial \psi (\vec r,t)}{ \partial t} = [- \frac{ \hbar^2 }{2 m} \nabla^2 + 
\frac{m \omega_{0}^2 r^2}{2} + V_{H}(r) ] \psi (\vec r,t),
\end{equation}

 where $ V_{H} (\vec r) $ is the self-consistent  Hartree potential,

\begin{equation}
V_{H}(\vec r) = \frac{4 \pi a \hbar^2 }{m} |\psi (\vec r,t)|^2 -
 u \int d^3r^{\prime} \frac{| \psi ( \vec r^{\prime},t)|^2}{ |\vec r - \vec r^{\prime}| }
\end{equation}
The normalization condition for $ \psi $ is $ \int | \psi (\vec r,t) |^2 d^3 r = N $, $ N $ is 
the total number of particles in the condensed state. The original Gross-Pitaevskii 
equation can be obtained by putting $ u = 0 $.

One can write down the Lagrangian density corresponding to this system as follows,
\begin{equation}\label{lag}
{\cal L}  =  \frac{i \hbar }{2} ( \psi \frac{\partial \psi^{*}}{\partial t} -
 \psi^{*} \frac{\partial \psi }{\partial t}) + \frac{ \hbar^2 }{2m} |\nabla \psi |^2 
+ \frac{m \omega^2 r^2}{2} |\psi |^2  +  \frac{2\pi a \hbar^2 }{m} |\psi|^4 - 
\frac{u}{2} |\psi|^2 \int d^3r^{\prime} \frac{| \psi ( \vec r^{\prime},t)|^2}
{ |\vec r - \vec r^{\prime}| },
\end{equation}
where * denotes the complex conjugation.
The non-linear Schroedinger equation can be obtained from a minimization of the action,
$ I = \int {\cal L} d^3 r dt $. The BEC of charged Bosons \cite{wadati} confined 
in an ion trap can be described by the above mentioned Lagrangian if we
set $- u = e^2 $, where $ e $ is the electronic charge.

To calculate the excitations spectrum of an atomic BEC with gravity-like interaction, we 
will use the time-dependent variational method.  This technique has been first used
to calculate the low-lying excitations spectrum of a harmonically trapped atomic BEC
in Ref. \cite{gar}.
The result obtained from the variational method matches with Stringari's \cite{strin}
result within the sum-rule approach.

In Ref. \cite{gio}, it is shown that the oscillation frequencies obtained
from the exact ground state and a Gaussian ansatz are in good agreement.
In this system, a Gaussian ansatz is also a good variational wave function.
In order to obtain the evolution of the condensate we assume the following variational
 wave function,
\begin{equation}\label{wave}
\psi(\rho, z, t) = C(t) e^{-\frac{1}{2}[\alpha (t) \rho^2 + \beta (t) z^2]},
\end{equation}
 where $ C(t) $ is the normalization constant which can be determined from the 
normalization
condition.  $ \rho $ and $ z $ are the variables in units of $ \Lambda $, 
where $ \Lambda = \sqrt{ \hbar / m \omega_g } = \hbar^2 /muN $ is the 
length scale in this system ( similar to the harmonic osscilator length ) 
when the harmonic trap is absent and
$ \omega_g = m u^2 N^2 /\hbar^3 $ is the gravitational frequency.
$ \vec \rho $ is the two dimensional vector.
$ \alpha (t) = 1/\alpha_{1}^2 + i \alpha_2 $ and $ \beta (t) = 1/\beta_{1}^2 + i \beta_2 $ are the 
dimensionless time dependent parameter. $ \alpha_1 $ and $ \beta_1 $ are the condensate widths 
in $ x-y $ plane and along the $ z $ direction respectively. The Gaussian variational wave 
function is an exact ground state when the two two-body interatomic interaction is absent.

Substituting (\ref{wave}) into (\ref{lag}) and integrating the Lagrangian density over the 
space co-ordinates. We get the following Lagrangian,
\begin{equation} \label{lag1}
L  =  \frac{S Nu}{2a}[ (\alpha_{1}^2 \dot \alpha_2 + \frac{1}{2} \beta_{1}^2 \dot \beta_2 ) 
- ( \frac{1}{\alpha_{1}^2} + \alpha_{1}^2 \alpha_{2}^2 )\\ \nonumber  -  
\frac{1}{2}(\frac{1}{\beta_{1}^2}
+ \beta_{1}^2 \beta_{2}^2 )  -  \sqrt {\frac{2}{\pi}}(\frac{S}{ \alpha_{1}^2 \beta_1 } 
- \frac{ F[\frac{1}{2}, 1; \frac{3}{2}; (1-\frac{\alpha_{1}^2}{\beta_{1}^2})]}{ \beta_1})].
\end{equation}
where $ S = \frac{N a}{ \Lambda } =  \frac{ u m a N^2 }{ \hbar^2} $ is a 
dimensionless 
scattering parameter similar to the scattering parameter $ P = \frac{ N a }{a_0} $
 \cite{gar} for an ordinary atomic BEC. Here, $ a_0 $ is the harmonic osscilator
length. This $ S $ can be positive or negative depending on the sign of the
scattering length $ a $. The scattering parameter $ S $ can also be written as
$ S = (528 \pi^2/105) \tilde {I} ( N a / \lambda_L )^2 $, where $ \tilde {I} = I/I_0 $ and 
$ I_0 = (48 \pi /7) (c \hbar^2 \epsilon_{0}^2/m \alpha_0^2 ) a $ is the threshold laser intensity to
create a self-bound condensate \cite{kuri}.
For a given intensity $ \tilde {I} = 1.5 $, the realistic, stable and self-trapped system
( for sodium atoms ) contains 40 to 400 atoms \cite{kuri} and the corresponding range of $ S $ 
is 1 to 100. This range can be alter by changing the scattering length $ a $ \cite{in}. 
$ F[\frac{1}{2}, 1; \frac{3}{2}; (1-\frac{\alpha_{1}^2}{\beta_{1}^2})] $ is the Hypergeometric 
function. The last term in the Eq. (\ref{lag1}) is the mean-field energy of the gravity-like
potential. 
We are interested to find out the excitation spectrum of a self-bound Bose gas as well as 
in the TF-G and G regimes. 
We have set $ V_{ext} = 0 $ because the system is stable even in the absence of an external 
trap potential. 

The energy functional in terms of the variational parameter $ \alpha $ in 
an isotropic system is
\begin{equation}
E = \frac{Nu S }{2a}[ \frac{3}{2 \alpha^2} +
    \sqrt {\frac{2}{\pi}}( \frac{S}{\alpha^3} - \frac{1}{\alpha})],
\end{equation}
By minimizing the energy functional with respect to $ \alpha $, one can get 
 the equilibrium point $ w $ which is given by,

\begin{equation}\label{width}
w = \frac{3}{2} \sqrt{\frac{\pi}{2}} [ 1 + \sqrt{( 1 + \frac{8 S}{3 \pi})}].
\end{equation}

The sound velocity is $ c_s^2 = \mu /m $, 
where $ \mu = u S/2a[ 3/2 w^2 + 
   2 \sqrt{2/\pi}( S/w^3 - 1/w)] $ is the chemical 
potential. The sound velocity $ c_s $ vs the dimensional scattering parameter $ S $ is shown
in Fig.1. 

Using the Euler-Lagrange equation, the time evolution of the 
widths are,

\begin{equation}\label{equ1}
\ddot{\alpha_1} = \frac{1}{\alpha_{1}^3} + 
\sqrt {\frac{2}{\pi}} (\frac{S}{\alpha_{1}^3 \beta_{1}}
+ \frac{ F_{\alpha_1}[\alpha_1, \beta_1]}{2}),
\end{equation}

\begin{equation}\label{equ2}
\ddot{\beta_1} = \frac{1}{\beta_{1}^3} 
+ \sqrt {\frac{2}{\pi}} (\frac{S}{\alpha_{1}^2 \beta_{1}^2}
+  F_{\beta_1}[\alpha_1, \beta_1]).
\end{equation}

$ F_{\alpha_1}[\alpha_1, \beta_1] $ is the derivative of 
$ F[\frac{1}{2}, 1; \frac{3}{2}, (1 - \frac{\alpha_{1}^2}{\beta_{1}^2})]/\beta_1 $
with respect to $ \alpha_1 $. 
Similarly, $ F_{\beta_1}[\alpha_1, \beta_1] $ is the derivative of
$ F[\frac{1}{2}, 1; \frac{3}{2}, (1 - \frac{\alpha_{1}^2}{\beta_{1}^2})]/\beta_1 $ 
with respect to $ \beta_1 $. The exact form of 
$ F_{\alpha_1}[\alpha_1, \beta_1] $ and 
$ F_{\beta_1}[\alpha_1, \beta_1] $ are given in the Appendix A. 

We are interested in the low-energy excitations of a gravity-like
self-bound Bose condensate.
The low-energy excitations of the condensate corresponds to the small oscillations of
the state around the equilibrium widths $ \alpha_{10} $ and $ \beta_{10} $. 
Therefore, we expand around the time dependent variational parameters around
the equilibrium widths in the following way,
 $ \alpha_1 = \alpha_{10} + \delta \alpha_1 $ and
 $ \beta_1 = \beta_{10} + \delta \beta_1 $. 

The time evolution of the widths around the equilibrium points are

\begin{equation}\label{fluc1}
\ddot {\delta \alpha_1}  = - (\frac{3}{\alpha_{10}^4} + 3 \sqrt {\frac{2}{\pi}}
\frac{S}{\alpha_{10}^4 \beta_{10}}) \delta \alpha_1 - \sqrt
{\frac{2}{\pi}} \frac{S}{\alpha_{10}^3 \beta_{10}^2} \delta \beta_1
 +\frac{1}{2}\sqrt{\frac{2}{\pi}} F_{\alpha_1}[ \alpha_{10}, \beta_{10},
 \delta \alpha_1, \delta
\beta_1 ],
\end{equation}

\begin{equation}\label{fluc2}
\ddot {\delta \beta_1}  = - 2 \sqrt {\frac{2}{\pi}} \frac{S}{\alpha_{10}^3 \beta_{10}^2} \delta
\alpha_1 - (\frac{3}{\beta_{10}^4} + 2 \sqrt {\frac{2}{\pi}} \frac{S}{\alpha_{10}^2
\beta_{10}^3}) \delta \beta_1  + \sqrt{\frac{2}{\pi}} F_{\beta_1}[\alpha_{10}, \beta_{10}, 
\delta \alpha_1, \delta \beta_1 ].
\end{equation}

$ F_{\alpha_1}[\alpha_{10}, \beta_{10}, \delta \alpha_1, \delta\beta_1 ] $ is the first order
fluctuations around the equilibrium points of $ F_{\alpha_1}[\alpha_1, \beta_1] $.
Similarly, $ F_{\beta_1}[\alpha_{10}, \beta_{10}, \delta \alpha_1, \delta\beta_1 ] $ is the first
order fluctuations around the 
equilibrium points of $ F_{\beta_1}[\alpha_1, \beta_1] $.
The exact form of $ F_{\alpha_1}[\alpha_{10}, \beta_{10}, \delta \alpha_1, \delta\beta_1] $ 
and $ F_{\beta_1}[\alpha_{10}, \beta_{10}, \delta \alpha_1, \delta\beta_1] $  are given in the
Appendix A.
 
We are looking for the solutions of $ e^{i \omega t } $ type. First we solve for 
these two equation in terms of $ \alpha_{10}, \beta_{10} $ and later we set  
$ \alpha_{10} = \beta_{10} = w $. For isotropic system, the 
excitations spectrum are

\begin{equation}
\frac{\omega_{+}^2}{\omega_{g}^2}  =  \frac{3}{ w^4} 
+  \sqrt{\frac{2}{\pi}} ( \frac{4 S}{w^5} -
\frac{2}{3 w^3}), 
\end{equation}

\begin{equation}
\frac{\omega_{-}^2}{\omega_{g}^2}  =  \frac{3}{ w^4}            
+  \sqrt{\frac{2}{\pi}} ( \frac{S}{w^5} -   
\frac{1}{15 w^3}),      
\end{equation}

where $ \omega_g = \frac{m u^2 N^2}{\hbar^3} $ is the gravitational frequency.
The $ \omega_{+} $ and $ \omega_{-} $ vs. the dimensionless scattering parameter $ S $
are shown in Fig2.
When $ S $ is small, the gravity-like $ 1/r $ attractive interaction
dominates over the repulsive pseudopotential. In this limit, the system
becomes more compressible. In other words, the system becomes less resistant
to density changes. So one would expect that monopole mode lies below
quadrupole mode.
From the Fig.2, we identify that the upper branch of the excitation spectrum
 is quadrupole mode
( $\omega_{-} = \omega_Q $ ) and lower branch is monopole mode 
( $\omega_{+} =  \omega_M $ ).
 The monopole and quadrupole
modes spectrum in Fig.2 matches very well with the spectrum obtained within the
sum rule approach \cite{gio}.
When $ S = -1.169 $, $ \omega_M $ starts decreasing as shown in Fig.2. 
At $ S_c = - 1.179 $, the system collapses. The value of $ S_c $ can also be 
obtained from the Eq. (\ref{width}). 
For large value of $ S $, the monopole mode lies above the quadrupole mode
because the repulsive pseudopotential start dominates over 
the attractive long-range interaction. At $ S = 17.5 $, there is a crossing
between these two modes which is shown in inset of Fig.2. Interestingly, this crossing
of these two modes is also obtained from the time-dependent variational method.

{ \bf TF-G regime: }
For large $s$-wave scattering length, the kinetic and the trap potential energy 
can be neglected. The
gravity-like potential is balanced by the  $ s $-wave interaction strength. 
The total ground state energy is $ E_{0} = - 0.9648 N^2 ( u/ \Lambda_{J} ) $,
where $ \Lambda_{J} = 2 \pi \sqrt{a \hbar^2 / m u} $ is the Jeans wavelength which
is the shortest wavelength to keep stable condensed state. The ground state 
energy per particle varies as $ N $. The sound velocity is
$ c_s^2 = \mu /m = 0.307106 (u \sqrt{S})/m ) $. So the sound velocity
 $ c_s $ varies as $ N^{1/2} $ whereas $ c_s \sim N^{1/5} $ for 
an ordinary atomic BEC in the TF approximation \cite{baym}. 
In this regime, we neglect the contribution of the kinetic energy term
in Eqs. (\ref{equ1}) and (\ref{equ2}) and 
we find that the monopole and quadrupole frequencies  are
 $ \omega_M = 0.319951 \omega_g S^{-3/4} $ 
and $ \omega_Q = 0.202355  \omega_g S^{-3/4} $.
In this regime, the monopole and quadrupole frequencies are obtained by solving the
hydrodynamic equations numerically in \cite{gio}. The monopole and quadrupole frequencies
obtained from the variational approach are similar to the exact numerical values. 
For an ordinary atomic BEC in the TF regime, the $ \omega_M $ and $ \omega_Q $ are independent
of the scattering length $ a $ \cite{gar,strin}. But, in this system, it still depends on 
the scattering length $ a $.
Here, the ratio $ \omega_M /\omega_Q $ is $ \sqrt{5/2} = 1.58114 $. Remarkably, 
this ratio is identical to the result of \cite{gio} which is 
obtained within the sum-rule approach. This is also true for trapped atomic BEC without
gravity-like interaction in the large N limit. It was first pointed out in \cite{gio,strin}.

{ \bf G regime: }  In this regime, gravity-like potential is balanced by the kinetic energy. 
The trap potential and $s$-wave interaction can be neglected. This is analog of Boson-star 
(non-relativistic) \cite{rufi}. The total ground state energy is 
$ E_{0} = - (N /19) \hbar \omega_g $. Using the uncertainty relation, we estimate the total ground
 state energy is $ E_{0} = - (N /16) \hbar \omega_g $ which is very close to the energy obtained 
from the variational approach. So our variational ansatz for wave function is good in 
this regime also. 
The ground state energy per particle varies as $ N^2 $. The sound velocity is
$ c_s^2 = 0.159155 u S /m  $. So the sound velocity
 $ c_s $ varies as $ N $.
In this regime, neglecting
the contribution of the $s$-wave interaction potential in Eqs. (\ref{equ1}) and 
(\ref{equ2}), we get
the monopole and quadrupole modes are $ \omega_M = 0.0707355 \omega_g   $ and
$ \omega_Q = 0.118363 \omega_g $. 
These  frequencies are very close to the frequencies obtained within
the sum-rule approach \cite{gio}.
Here, $ \omega_M / \omega_Q = \sqrt{5/14} = 0.597615 $.
This ratio is also identical (up to five decimal) with that of \cite{gio}, 
which is obtained within the sum-rule approach. 

\section{vortices of a gravitationally self-bound Bose condensate state}

We consider a gravitationally self-bound Bose condensate state with 
a vortex along the $ z$-axis. The experimental realization of a 
vortex state would be a direct signature of macroscopic phase coherence of
this new atomic BEC with a attractive long-range interaction. 
One can use the time-dependent variational approach to describe the vortex
 state. In the previous section, we have 
explicitly shown that the monopole  and quadrupole mode frequencies obtained 
by using the Gaussian ansatz coincides with the numerical results. So it is
natural choice to assume a variational wave function 
of a self-bound BEC state with a vortex along the $ z $-axis is,
\begin{equation}\label{wave1}
 \psi_{q}(\vec r, t) = C_{q}(t) \rho^{q} e^{i q \phi }
 e^{-\frac{r^2}{2}(\frac{1}{\alpha^2(t)}+ i \beta (t))},
\end{equation}
where $ q $ is the vortex quantum number and $ C_q(t) $ is the
normalization constant. Also, $ \rho^2 = x^2 + y^2 $,
$ r^2 = x^2 + y^2 + z^2 $ and $ \phi = tan^{-1}(y/x) $.
For simplicity, we consider only $ q = 1 $ and $ q = 2 $.
By following the same procedure of the previous section,
one would obtain the effective Lagrangian which is
given by

\begin{equation}
 L = \frac{NuS}{2a} [(q+\frac{3}{2}) \alpha^2 \dot {\beta} 
 - (q+\frac{3}{2})(\frac{1}{\alpha^2 } + \alpha^2 \beta^2)
 - \sqrt{\frac{2}{\pi}} (\frac{g_q S }{\alpha^3} - \frac{c_q}{\alpha})],
\end{equation}
 where $ g_q = \frac{(2q)!}{2^{2q}(q!)^2}$, $ c_1 = 23/30 $ and
 $ c_2 = 37/56 $.

The energy functional of the vortex state in terms of the variational
parameter $ \alpha $ is

\begin{equation}
E_q = \frac{NuS}{2a}[ (q+\frac{3}{2})(\frac{1}{\alpha^2}) 
 + \sqrt{\frac{2}{\pi}} (\frac{g_q S }{\alpha^3} - \frac{c_q}{\alpha})]
\end{equation}

By minimizing the energy with respect to the variational 
parameter $ \alpha $, one could obtain the equilibrium width $ w_q $ 
which is given by
\begin{equation}\label{widthq}
w_q = \frac{ \sqrt{2 \pi} (q+ \frac{3}{2}) + \sqrt{ 2 \pi (q + \frac{3}{2})^2
+  12 g_q S c_q}}{2 c_q},
\end{equation}
where $ g_q $ and $ c_q $ are given above. 
The system collapses when $ S_c = - 8.53694 $ for $ q = 1 $ and 
$ S_c = - 25.8875 $ for $ q = 2 $. This critical value $ S_c $ is
increasing with increasing of the number of
vorticity. The expectation value
of the square of the system radius is 
$ I_q = \sqrt{<r^2>} = \sqrt{ N (q+3/2)} w_q $.
The energy functional satisfies the stability condition,
$ \frac{\partial^2 E_q}{\partial \alpha^2}|_{\alpha = w_q} > 0 $. 
 
The superfluid coherence length, $ \xi $, is a distance
over which  the condensate wave function can heal. In the case
of a vortex, it corresponds to the distance over which the
wave function increases from zero, on the vortex axis, to 
the bulk density. It can be calculated by equating the kinetic
energy to the interaction energies. The kinetic energy term can not
be neglected even for large $ S $, since it determines the 
structure of the vortex core. The system exhibits superfluid 
properties if the coherence length  is small compared to 
the size of the condensed state, otherwise it will be less
prominent to observe the superfluid properties.
Equating the kinetic energy to the interaction energies,

\begin{equation}
\frac{\hbar^2}{2m \xi^2} = \frac{4 \pi a N \hbar^2}{m R^3} - 
\frac{u N}{\xi},
\end{equation}
where $ R =  a N \sqrt{F} / S $ is the radius of the condensed state at $ S $
and,
\begin{equation}\label{width1}
 F = \frac{3}{2 w^2} + \sqrt{\frac{8}{\pi}}( \frac{S}{w^3} - \frac{1}{w}) .
\end{equation}
$ w $ is given in the Eq. (\ref{width}). After rescaling the above equation,
we get a quadratic equation of $ \xi $ whose solution is,
\begin{equation}\label{coh}
\frac{ \xi }{ R } = \frac{ F + \sqrt{ F^2 + 8 S \sqrt{F}}}{8 \pi S }.
\end{equation}
The superfluid coherence length $ \xi /R $ vs. $ S $ is shown
in Fig.3 for a wide range of $ S $.
The vortex state play an important role in characterizing
the superfluid properties of Bose system. 
The critical angular frequency required to produce a vortex state is \cite{dalfovo}
\begin{equation}
\Omega_{q} = \frac{(E_q - E_0)}{N \hbar q },
\end{equation}

where $ E_q $ is the energy of a vortex states with
vortex quantum number $ q $ and $ E_0 $ is the energy with
no vortex. The critical angular frequency $ \Omega_{1} $
 vs. the dimensionless scattering parameter $ S $ is
shown in Fig.4. The critical angular 
frequency decreases with increasing $ S $ (or $ N $).
For attractive interaction, $ \Omega_q $ increases with 
increasing of $ S $.
This is also true for an ordinary BEC in the TF regime \cite{dalfovo}. 

The monopole mode frequency for the vortex state with
a vortex number $ q $ is

\begin{equation}
\frac{\omega_q^2}{\omega_g^2} =
 [\frac{3}{w_q^4} + \sqrt{\frac{2}{\pi}}\frac{1}{(q+\frac{3}{2})}
(6 \frac{ S g_q}{w_q^5} - \frac{c_q}{w_q^3})],
\end{equation} 
where $ w_q $ is given in the Eq. (\ref{widthq}).

{ \bf TF-G regime:} For large $ s $-wave scattering length, kinetic 
energy can be neglected. 
The superfluid coherence length can be obtain from the Eq. (\ref{coh}).
When $ S $ is large, one would gets $ F = 0.6142 S^{-1/2} $ from
Eq. (\ref{width1}). Then, $ \xi/ R = 0.1765  S^{-5/8} $. When is $ S $ is very
large, coherence length is very small compared to the size of the
system. The TF-G regime should exhibit the superfluid properties. 
The mean size of the 
condensate with $ q = 1 $ and $ q = 2 $ are $ I_1 = 2.21163 \sqrt{S}  $ and
$ I_2 = 2.4412 \sqrt{S} $ respectively.
The size of a condensate state with vortices increases with
the number of vorticity. 
The critical angular frequencies for 
$ q =1 $ and $ q =2 $ are $ \Omega_{1} = \omega_g 0.0077 S^{-1/2} $
and $ \Omega_{2} = \omega_g 0.0094 S^{-1/2} $ respectively. 
The monopole mode frequencies for one and two vortices are
$ \omega_1 = 0.299   \omega_g S^{-3/4} $ and 
$ \omega_2 = 0.26 \omega_g  S^{-3/4} $,
respectively. These two monopole mode frequencies are less than
the monopole mode frequency of the vortex free condensate.
So, in the TF-G limit, monopole mode frequency of the condensate decreases
due to the presence of the vortex.  The monopole mode frequency
for an ordinary atomic BEC in the TF regime  is independent of
the vortex \cite{salas}.

{\bf G regime:} In this regime, we neglect the contribution from
the $s$-wave interaction energy. The superfluid coherence length
 $ \xi  $ can be obtained by equating the kinetic energy to the 
gravity-like interaction energy, $ \hbar^2/2m \xi^2 = uN/ \xi $,
which gives $ \xi = 0.8862 R_g $ which is almost equal to
the radius $ (R_g) $ of the condensate state in this regime. As 
we know that if the coherence length is compareable to the radius 
of the condensate state, it is difficult to exhibit the superfluid
properties.
In this regime, $ I_1 = 12.92 $ and
$ I_2 = 24.84 $. Here, $ I_2 >> I_1 $. The size of a condensate state 
(with vortices) expanding abruptly with increase of number of the vorticity.
For example, $ I_1/I_0 \sim 2.8 $ and $ I_2/I_0 \sim 5.39 $,
where $ I_0 = 4.604 $ is the mean size of the vortex free condensate.
The critical angular frequencies for $ q =1 $ and $ q =2 $ are 
$ \Omega_{1} =  0.0343 \omega_g $ and 
$ \Omega_{2} = 0.0215  \omega_g $, respectively. 
Here, $ \Omega_{2} <  \Omega_{1} $. In this regime, the condensate state 
with vortex of $ q = 2 $ is unbounded because $ |\mu_2 | /\hbar < 2 \Omega_{2} $,
where $ \mu_2 = - 0.0297 u S/a $ is the chemical potential in the rotating frame.
So the vortex of $ q \geq 2 $ can not be created in this regime
unless there is an additional repulsive potential.
Note that although there is an indication of instability of vortex with $ q = 2 $,
this may be just an artifact of the variational approach. 
The monopole mode frequency for $ q = 1 $ is $ \omega_1 = 0.0149 \omega_g $.    
The $ \omega_1 $ is also less than the monopole mode 
frequency $ \omega_M $ in the vortex free condensate.  
 
It should be noted that the vortex has two different length scales, condensate
radius and core radius, whereas the trial wave function (\ref{wave1}) has only one variational
length scale $(\alpha)$. 
In this variational approach, the various numerical values computed for the energies of the
two regimes and for the collapse are just indicative but more accurate values can be obtained
by other rigorous methods.

\section{summary and conclusions}
In  this paper,  we have derived an analytic expressions  of the  monopole  and quadrupole excitation 
frequencies  of a  self-bound Bose  gas  induced  by the electromagnetic wave  for a 
wide range of
the  dimensionless  scattering parameter $ S $. Later, we consider the two new regimes, namely, TF-G
and G regimes. In these regimes, we have calculated the lower bound of the ground state energy,
sound velocity, monopole and quadrupole mode frequencies. Our results are in excellent agreement with 
the result \cite{gio} obtained by using the sum-rule approach. 
Interestingly, the ratio $ \omega_M / \omega_Q $ is identical 
(in both the regimes)
with that of \cite{gio} which is obtained within the sum-rule approach.
In the TF regime of an ordinary atomic BEC, the monopole and quadrupole
mode frequencies are independent of the scattering length $ a $.
On the other hand, in the TF-G regime, the monopole and quadrupole mode
frequencies depends on the scattering length $ a $. The local sound velocity
$ c_s $ varies as $ N^{1/2} $ in the TF-G regime, whereas 
$ c_s \sim N^{1/5} $ for an ordinary atomic BEC in the TF regime.
For harmonic trapped Bose system, the excitation frequencies are
determined by the oscillator  frequency of the trap potential.
But, in this system, the monopole and quadrupole mode frequencies
are fixed by  the gravitational frequency $ \omega_g $.

In section III, based on the time-dependent variational method
and simple ansatz for the wave function (equ. \ref{wave1} ), we have studied 
a rotating gravity-like self bound BEC states with vortices
along the $ z $-axis.
We derived an analytic expressions for the coherence length and 
the critical angular frequency to create a vortex in the
condensed state. We found that the coherence length in the
TF-G regime is very very small compared to the radius of the
system. On the other hand, the coherence length in the
G-regime is compareable to the radius of the system. 
We could say that the TF-G regime should exhibit
the superfluid properties more prominently than the G-regime.
In the TF regime of an ordinary atomic BEC, the monopole mode frequency 
of the condensate does not change due to the presence of the 
vortex. But, the monopole mode frequency in the TF-G regime 
as well as in the G regime of 
this new BEC decreases due to the presence of the vortex.

I would like to thank G. Baskaran and Pijush K. Ghosh for valuable comments.

\begin{appendix}

\section{}
Here, we shall give the exact form of $ F[\alpha_1, \beta_1] = F[\frac{1}{2}, 1; \frac{3}{2}; 
 (1-\frac{\alpha_{1}^2}{\beta_{1}^2})]/\beta_{1} $, its derivative
with respect to $ \alpha_{1} $, $ \beta_1 $ and the first order
deviation from the equilibrium functions  
$ F_{\alpha_1}[\alpha_{10}, \beta_{10}] $ and $ F_{\beta_1}[\alpha_{10}, \beta_{10}]$.

\begin{equation}
F[\alpha_1, \beta_1] = \frac{1}{\beta_{1}}[ 1 + \frac{1}{3} ( 1 - 
\frac{\alpha_{1}^2}{\beta_{1}^2})
   + \frac{1}{5} ( 1 -2\frac{\alpha_{1}^2}{\beta_{1}^2}
   + \frac{\alpha_{1}^4}{\beta_{1}^4}) 
   +  \frac{1}{7} (1 -3\frac{\alpha_{1}^2}{\beta_{1}^2}
   + 3\frac{\alpha_{1}^4}{\beta_{1}^4} - \frac{\alpha_{1}^6}{\beta_{1}^6})
   + ........].
\end{equation}

The derivative of $ F[\alpha_1, \beta_1] $ with respect $ \alpha_1 $ is given by

\begin{equation}
F_{\alpha_1}[\alpha_1, \beta_1]  =  \frac{1}{\beta_{1}}[ 
-\frac{2}{3}\frac{\alpha_{1}}{\beta_{1}^2}
  + \frac{1}{5}(-4\frac{\alpha_{1}}{\beta_{1}^2} + 
  4 \frac{\alpha_{1}^3}{\beta_{1}^4} ) + 
  \frac{1}{7} (-6\frac{\alpha_{1}}{\beta_{1}^2} + 
  12 \frac{\alpha_{1}^3}{\beta_{1}^4} - 6 \frac{\alpha_{1}^5}{\beta_{1}^6})
  + ...........].
\end{equation}

Expanding $ F_{\alpha_1}[\alpha_1, \beta_1] $ around the equilibrium width
 $ \alpha_{10} $ and $ \beta_{10} $. The first order deviation from the 
equilibrium function $ F_{\alpha_1}[\alpha_{10}, \beta_{10}]$ is,
\begin{eqnarray}\label{b3}
F_{\alpha_1}[\alpha_{10}, \beta_{10}, \delta \alpha_{1}, \delta \beta_{1}] & = & 
\frac{1}{\beta_{10}}[  -\frac{2}{3}\frac{1}{\beta_{10}^2}
  +  \frac{1}{5}( - \frac{4}{\beta_{10}^2} + 12 
  \frac{\alpha_{10}^2}{\beta_{10}^4})
  +  \frac{1}{7} (-\frac{6}{\beta_{10}^2}  + 
  36 \frac{\alpha_{10}^2}{\beta_{10}^4} - 30 \frac{\alpha_{10}^4}{\beta_{10}^6})
   + ......] \delta \alpha_1 \\ \nonumber & - & 
   \frac{1}{\beta_{10}^2}[-\frac{2}{3}\frac{\alpha_{10}}{\beta_{10}^2}
  + \frac{1}{5}(-4\frac{\alpha_{10}}{\beta_{10}^2} +
  4 \frac{\alpha_{10}^3}{\beta_{10}^4} )  + 
  \frac{1}{7} (-6\frac{\alpha_{10}}{\beta_{10}^2} +
  12 \frac{\alpha_{10}^3}{\beta_{10}^4} - 6 \frac{\alpha_{10}^5}{\beta_{10}^6})
   +.........]\delta \beta_1
  \\ \nonumber & + &
  \frac{1}{\beta_{10}}[\frac{4}{3}\frac{\alpha_{10}}{\beta_{10}^3}
  + \frac{1}{5}(8\frac{\alpha_{10}}{\beta_{10}^3} - 16 \frac{\alpha_{10}^3}{\beta_{10}^5}) 
  + \frac{1}{7} (12 \frac{\alpha_{10}}{\beta_{10}^3} - 48 \frac{\alpha_{10}^3}{\beta_{10}^5}
  + 36 \frac{\alpha_{10}^5}{\beta_{10}^7}) + ........]\delta \beta_1.
\end{eqnarray}

The derivative of $ F[\alpha_1, \beta_1] $ with respect to $ \beta_1 $ is given by

\begin{eqnarray}
F_{\beta_1}[\alpha_1, \beta_1] & = & - \frac{1}{\beta_{1}^2} [ 1 + \frac{1}{3} ( 1 - 
\frac{\alpha_{1}^2}{\beta_{1}^2})
   + \frac{1}{5} ( 1 -2\frac{\alpha_{1}^2}{\beta_{1}^2}
   + \frac{\alpha_{1}^4}{\beta_{1}^4})
   +  \frac{1}{7} (1 -3\frac{\alpha_{1}^2}{\beta_{1}^2}
   + 3\frac{\alpha_{1}^4}{\beta_{1}^4} - \frac{\alpha_{1}^6}{\beta_{1}^6})
   + ........]
   \\ \nonumber & + &
   \frac{1}{\beta_1} [\frac{2}{3}\frac{\alpha_{1}^2}{\beta_{1}^3}
  + \frac{1}{5}(4 \frac{\alpha_{1}^2}{\beta_{1}^3} - 4 \frac{\alpha_{1}^4}{\beta_{1}^5})
  + \frac{1}{7} (6 \frac{\alpha_{1}^2}{\beta_{1}^3} - 12 \frac{\alpha_{1}^4}{\beta_{1}^5}
  + 6 \frac{\alpha_{1}^6}{\beta_{1}^7}) + ......].
\end{eqnarray}

Expanding $ F_{\beta_1}[\alpha_1, \beta_1] $ around the
 $ \alpha_{10} $ and $ \beta_{10} $. The first order deviation 
from the equilibrium function  $ F_{\beta_1}[\alpha_{10}, \beta_{10}]$
is the following,
 
\begin{eqnarray}\label{b5}
F_{\beta_1}[\alpha_{10}, \beta_{10}, \delta \alpha_{1}, \delta \beta_{1}] & = &
  - \frac{1}{\beta_{10}^2}[- \frac{2}{3}\frac{\alpha_{10}}{\beta_{10}^2}
  + \frac{1}{5}(- 4 \frac{\alpha_{10}}{\beta_{10}^2} 
  + 4 \frac{\alpha_{10}^3}{\beta_{10}^4}) +............] \delta \alpha_1
  \\ \nonumber & + &
  \frac{1}{\beta_{10}}[ \frac{4}{3}\frac{\alpha_{10}}{\beta_{10}^3}
  + \frac{1}{5}(8 \frac{\alpha_{10}}{\beta_{10}^3} - 16 \frac{\alpha_{10}^3}{\beta_{10}^5})
  + \frac{1}{7} (12 \frac{\alpha_{10}}{\beta_{10}^3} - 48 \frac{\alpha_{10}^3}{\beta_{10}^5}
  + 36 \frac{\alpha_{10}^5}{\beta_{10}^7}) + ......] \delta \alpha_1
  \\ \nonumber & + &
   \frac{2}{\beta_{10}^3}[ 1+ \frac{1}{3} ( 1 - \frac{\alpha_{10}^2}{\beta_{10}^2})
   + \frac{1}{5} ( 1 -2\frac{\alpha_{10}^2}{\beta_{10}^2}
   + \frac{\alpha_{10}^4}{\beta_{10}^4})
   +  \frac{1}{7} (1 -3\frac{\alpha_{10}^2}{\beta_{10}^2}
   + 3\frac{\alpha_{10}^4}{\beta_{10}^4} - \frac{\alpha_{10}^6}{\beta_{10}^6})
   + ........] \delta \beta_1
   \\ \nonumber & - &
   \frac{2}{\beta_{10}^2}[\frac{2}{3} \frac{\alpha_{10}^2}{\beta_{10}^3}
   +\frac{4}{5}(\frac{\alpha_{10}^2}{\beta_{10}^3} - 
   \frac{\alpha_{10}^4}{\beta_{10}^5}) + ........] \delta \beta_1
    \\ \nonumber & + &
   \frac{1}{\beta_{10}}[- 2 \frac{\alpha_{10}^2}{\beta_{10}^4}
   + \frac{1}{5}(-12 \frac{\alpha_{10}^2}{\beta_{10}^4} +20 \frac{\alpha_{10}^4}{\beta_{10}^6})
   + \frac{1}{7} (-18 \frac{\alpha_{10}^2}{\beta_{10}^4} + 60 \frac{\alpha_{10}^4}{\beta_{10}^6}
   - 42 \frac{\alpha_{10}^6}{\beta_{10}^8}) + ............] \delta \beta_1.
\end{eqnarray}

\end{appendix}

\begin{figure} 
\begin{center}
\epsfig{file=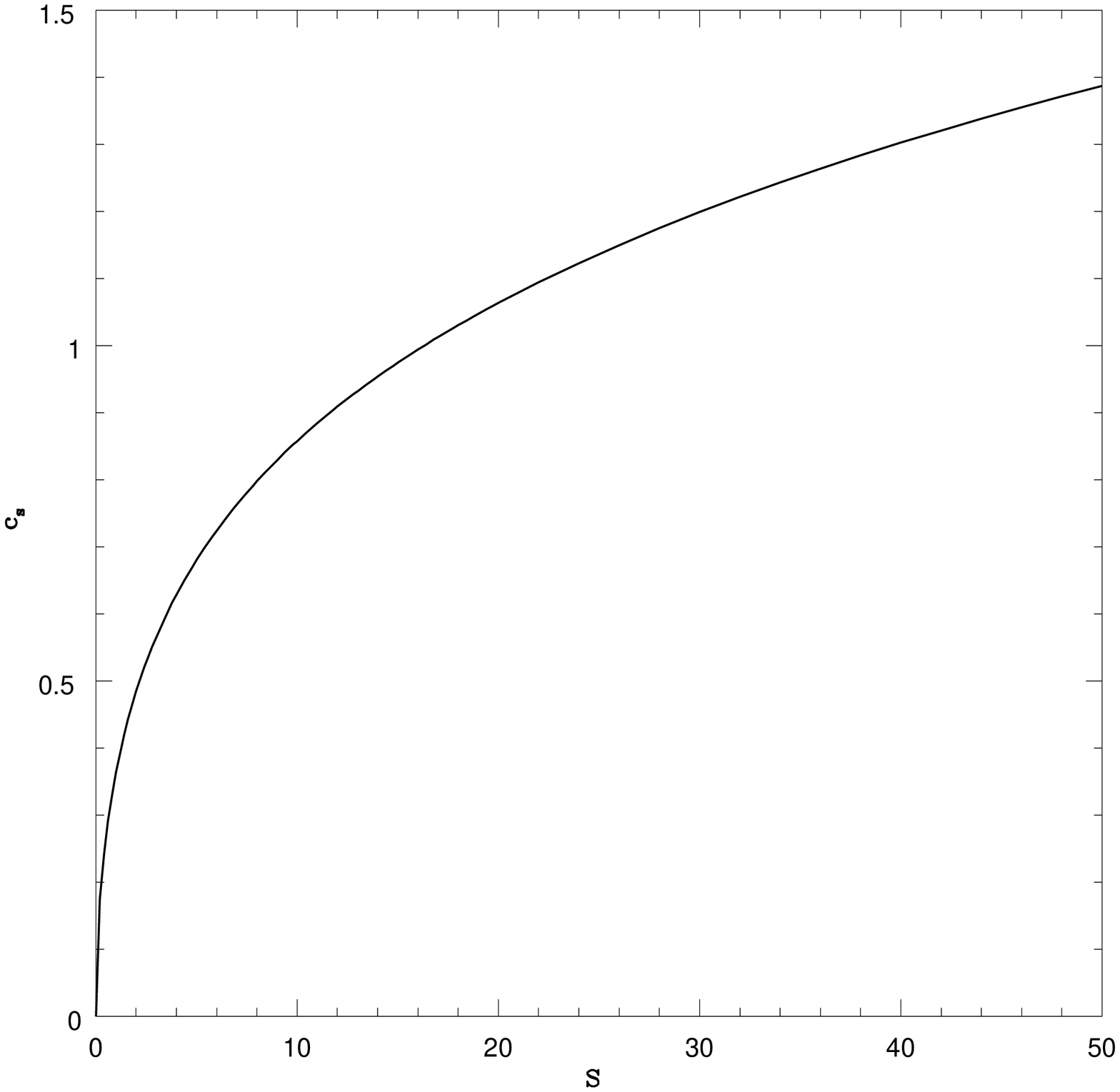, width= 10 cm,angle=0}
\vspace{.2 in} 
\begin{caption}
{ The sound velocity $ c_s $ as a function of the dimensionless
scattering parameter $ S $.}
\end{caption} 
\end{center}   
\label{fig1}
\end{figure}

\begin{figure}
\begin{center}
\epsfig{file=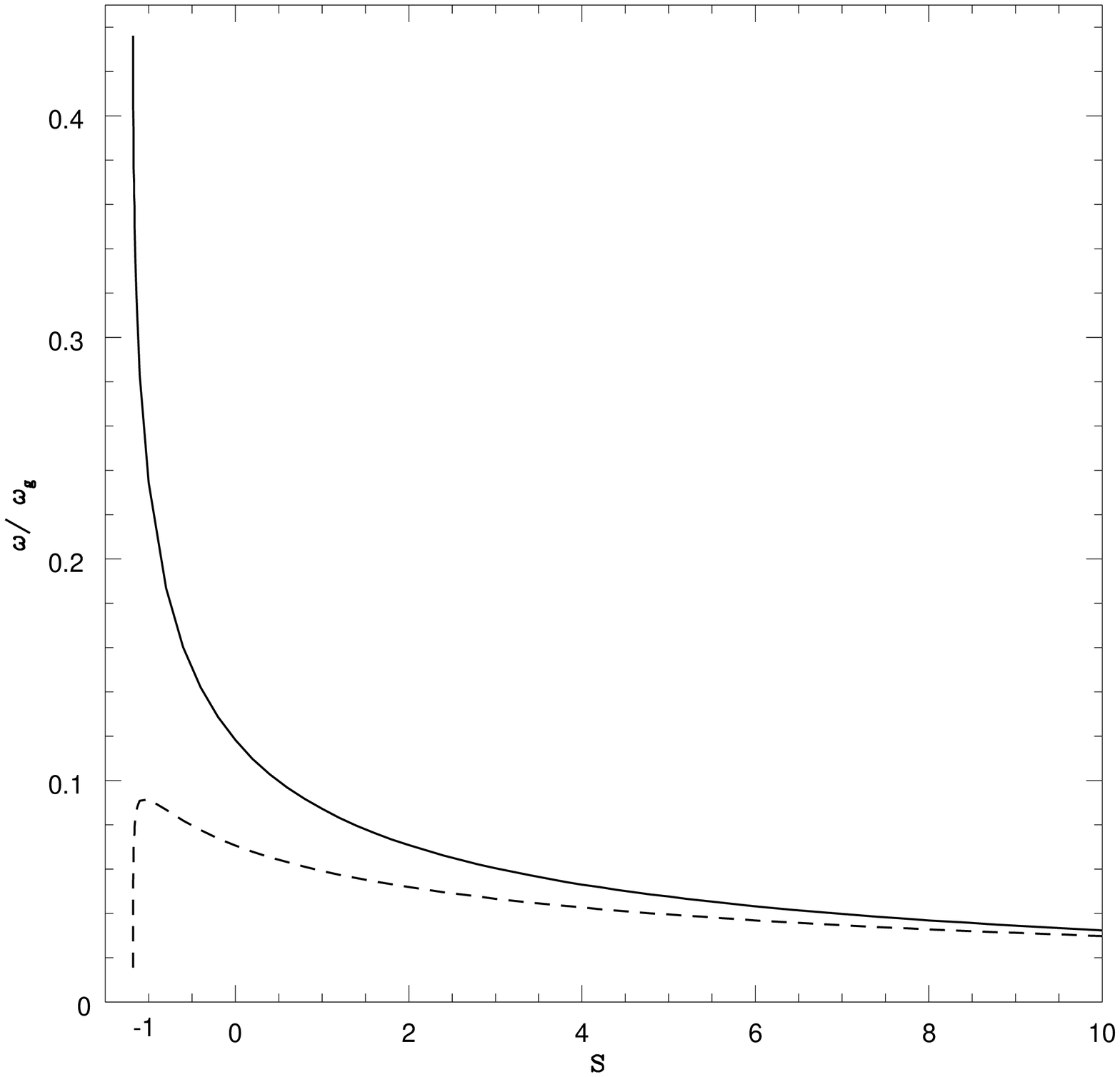, width= 10 cm,angle=0}
\begin{figure}
\vspace{-10cm}
\hspace{5cm}
\epsfig{file=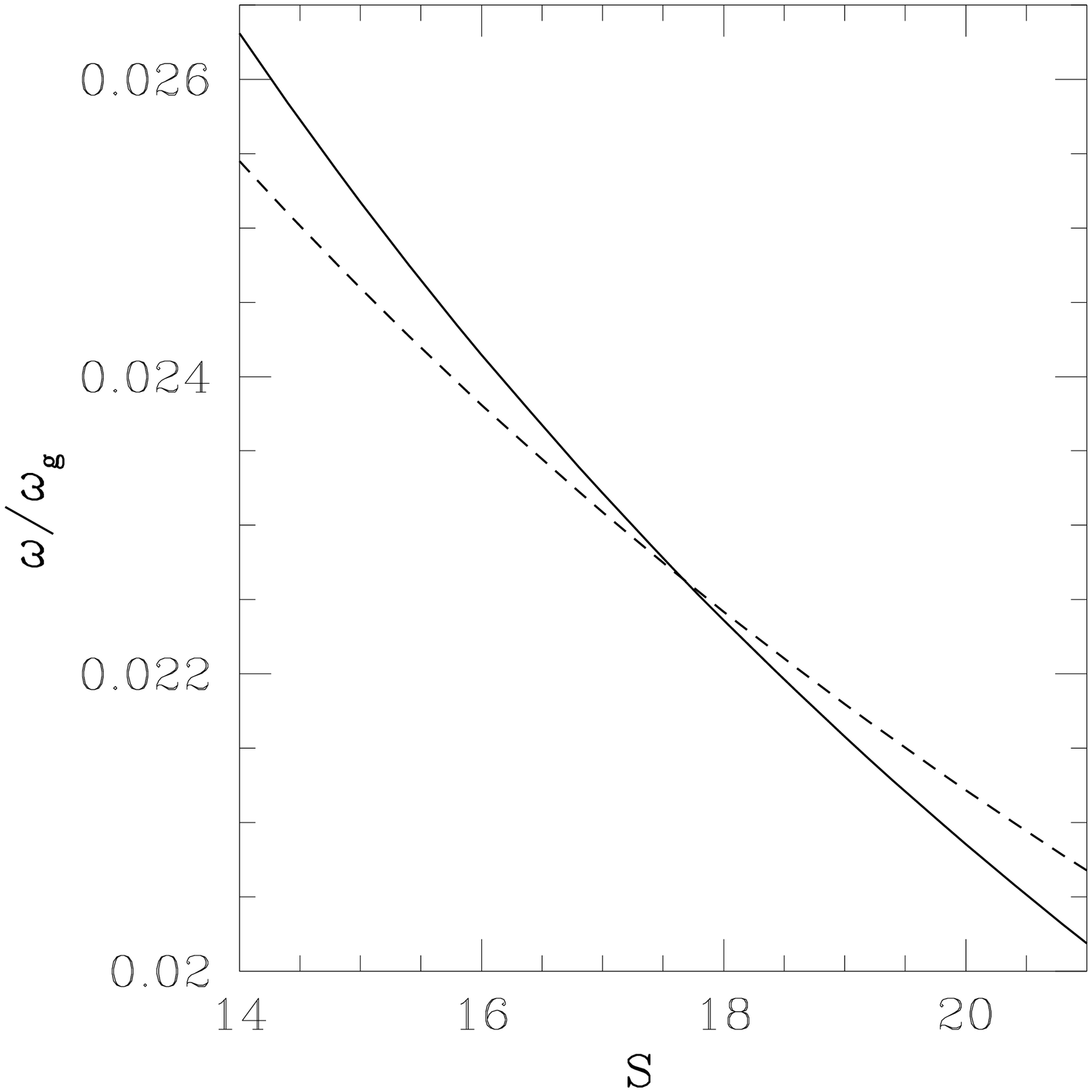, width= 4 cm,angle=0}
\end{figure}
\vspace{ 5 cm}
\begin{caption}
{Monopole and quadrupole mode frequencies vs. the scattering parameter
$ S $, for positive as well as negative values of $ a $. 
The solid and dashed lines corresponds to the  quadrupole and monopole
mode frequencies, respectively. The crossing between two modes
is shown in the inset of this figure.}
\end{caption}
\end{center}  
\label{fig2}
\end{figure}

\begin{figure}
\begin{center}
\epsfig{file=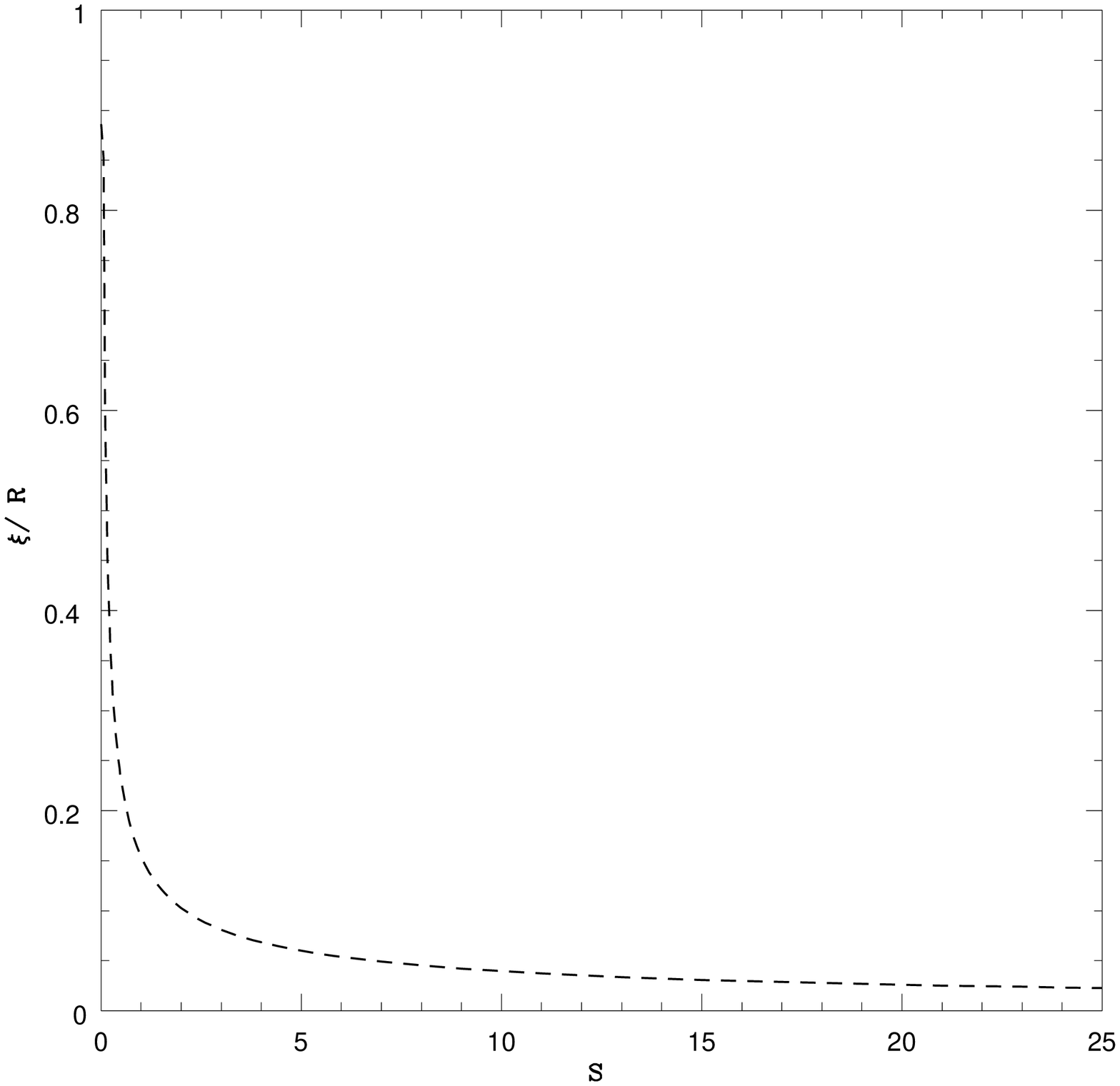, width= 10 cm,angle=0}
\vspace{.2 in}
\begin{caption}
{ The superfluid coherence length $ \xi $ as a function
  of the dimensionless scattering parameter $ S $. }
\end{caption}
\end{center}
\label{fig3}
\end{figure}

\begin{figure}
\begin{center}
\epsfig{file=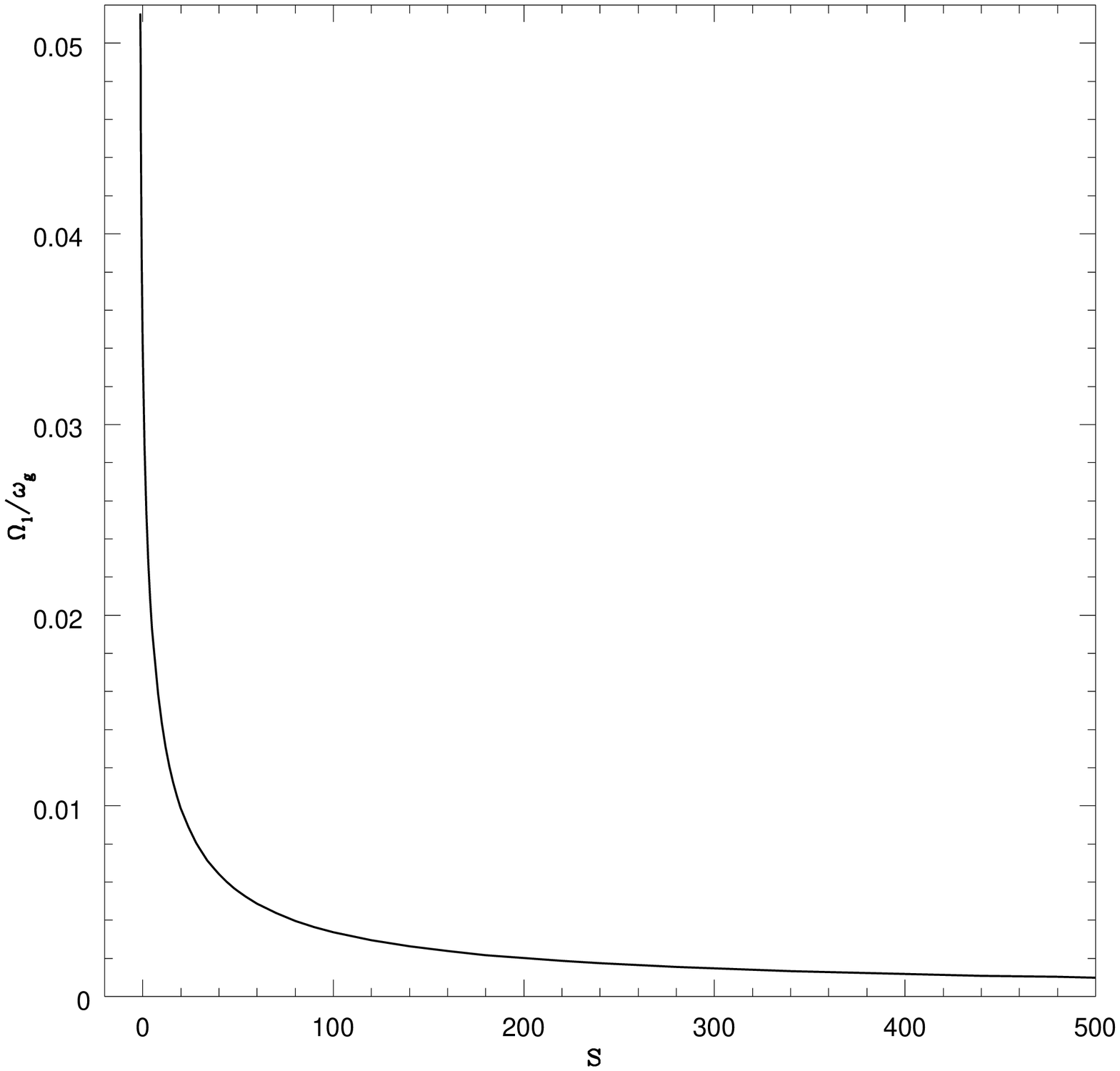, width= 10 cm,angle=0}
\vspace{.2 in} 
\begin{caption}
{ The critical angular frequency $ \Omega_{1} $ as a function
  of the dimensionless scattering parameter $ S $. }
\end{caption}
\end{center} 
\label{fig4} 
\end{figure}

\end{document}